\documentstyle[11pt,paspconf,epsf]{article}
 
\begin{document}
\title{Torus Construction}
\author{Monica Valluri \& David Merritt}
\affil{Department of Physics and Astronomy, Rutgers University, NJ}
 
\begin{abstract}
The maximally compact representation of a regular orbit is in terms of its
action-angle variables $({\bf J},\theta)$. Computing the map between a
trajectory's Cartesian coordinates and its action-angle variables is called
torus construction. This article reviews various approaches to torus
construction and their application to galactic dynamics.
\end{abstract}
 
\section{Introduction}
In systems with a single degree of freedom, constancy of the energy allows
the momentum variable $p$ to be written in terms of the coordinate variable
$q$ as $H(p,q) = E$, and the dependence of both variables on time follows
immediately from Hamilton's equations.  In general systems with $N \geq 2$
degrees of freedom (DOF), such a solution is generally not possible unless
the Hamilton-Jacobi equation is separable, in which case the separation
constants are isolating integrals of the motion. An isolating integral is a
conserved quantity that in some transformed coordinate system makes $\partial
H/\partial p_i = f(q_i)$, thus allowing the motion in $q_i$ to be reduced to
quadratures. Each isolating integral restricts the dimensionality of the
phase space region accessible to an orbit by one; if there are $N$ such
integrals, the orbit moves in a phase space of dimension $2N - N = N$, and
the motion is regular. The $N$-dimensional phase space region to which a
regular orbit is confined is topologically a torus (Figure 1). Orbits in
time-independent potentials may be either regular or chaotic, respecting 
a smaller number of integrals -- typically only the energy integral $E$.  
Although chaotic orbits are not, strictly speaking, confined to tori, 
numerical integrations suggest that many chaotic trajectories are effectively
regular, remaining confined for long periods of time to regions of phase
space much more restricted than the full energy hypersurface.
 
The most compact representation of a regular orbit is in terms of the
coordinates on the torus (Figure 1) -- the action-angle variables 
$({\bf J}, \theta)$. The process of determining the map $({\bf x,v}) 
\rightarrow
({\bf J},\theta)$ is referred to as {\it torus construction}. There are a number
of contexts in which it is useful to know the $({\bf J},\theta)$. One example is
the response of orbits to slow changes in the potential, which leave the
actions (${\bf J}$) unchanged.  Another is the behavior of weakly chaotic orbits,
which may be approximated as regular orbits that slowly diffuse from one
torus to another.  A third example is galaxy modeling, where regular orbits
are most efficiently represented and stored via the coordinates that define
their tori.
 
\sloppy
This article reviews techniques for mapping Cartesian coordinates into
action-angle variables in non-integrable potentials.  Two general approaches
to this problem have been developed.  Trajectory-following algorithms are
based on the quasi-periodicity of regular motion: Fourier decomposition of
the trajectory yields the fundamental frequencies on the torus as well as the
spectral amplitudes, which allow immediate construction of the map $\theta
\rightarrow {\bf x}$. Iterative approaches begin from some initial guess for
${\bf x}(\theta)$, which is then refined via Hamilton's equations with the
requirement that the $\theta_i$ increase linearly with time.  The two
approaches are often complementary, as discussed below.
 
\section{Regular Motion}
In certain special potentials, every orbit is regular; examples are the
Kepler and St\"ackel potentials. Motion in such potentials can be expressed
most simply by finding a canonical transformation to coordinates $({\bf p,q})$
for which the Hamiltonian is independent of ${\bf q}$, $H=H({\bf p})$; among all such
coordinates, one particularly simple choice is the action-angle variables
$(J_i,\theta_i)$, in terms of which the equations of motion are
\begin{eqnarray}
J_i & = & \hbox{constant,}\nonumber \\
\theta_i & = & \Omega_it + \theta_i^0,\ \ \ \ \Omega_i= {\partial
H\over\partial J_i}, \ \ \ i = 1,...,N
\end{eqnarray}
(Landau \& Lifshitz 1976; Goldstein 1980).  The trajectory 
${\bf x}({\bf J},\theta)$
is periodic in each of the angle variables $\theta_i$, which may be
restricted to the range $0<\theta_i\le 2\pi$. The $J_i$ define the
cross-sectional areas of the torus while the $\theta_i$ define position on
the torus (Figure 1).  These tori are sometimes called ``invariant'' since a
phase point that lies on a torus at any time will remain on it forever.
 
Most potentials are not integrable, but regular orbits may still exist;
indeed these are the orbits for which torus construction machinery is
designed.  One expects that for a regular orbit in a non-integrable
potential, a canonical transformation $({\bf x,v}) \rightarrow 
({\bf J},\theta)$
can be found such that
\begin{equation}
\dot J_i = 0,\ \ \ \ \dot \theta_i = \Omega_i, \ \ \ i = 1,...,N.
\end{equation} However there is no guarantee that the full Hamiltonian will
be expressible as a continuous function of the $J_i$.  In general, the map
$({\bf x}, {\bf v}) \rightarrow ({\bf J}, \theta)$ will be different for each orbit and
will not exist for those trajectories that do not respect $N$ isolating
integrals (although approximate maps, valid for some limited span of time,
may be derived for weakly chaotic trajectories).
 
\begin{figure}
\plotfiddle{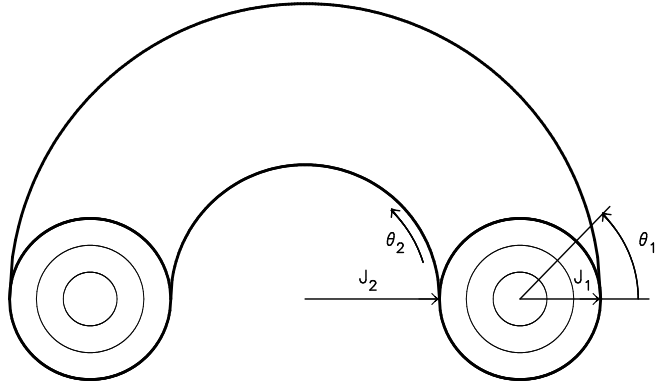}{6cm}{0}{75}{75}{-235}{-220}
\caption{Invariant torus defining the motion of a regular orbit in a
two-dimensional potential.  The torus is determined by the values of the
actions $J_1$ and $J_2$; the position of the trajectory on the torus is
defined by the angles $\theta_1$ and $\theta_2$, which increase linearly with
time, $\theta_i = \Omega_it + \theta_i^0$.  } \label{fig1}
\end{figure}
 
The uniform translation of a regular orbit on its torus implies that the
motion in any canonical coordinates $({\bf x}, {\bf v})$ is quasi-periodic:
\begin{eqnarray}
{\bf x}(t) & = &\sum_k {\bf X}_k({\bf J}) \exp\left[ i\left(l_k\Omega_1 + m_k\Omega_2 +
n_k\Omega_3\right)t\right], \nonumber \\
{\bf v}(t) & = &\sum_k{\bf V}_k({\bf J}) \exp\left[ i\left(l_k\Omega_1 + m_k\Omega_2 +
n_k\Omega_3\right)t\right],
\end{eqnarray}
with $(l_k,m_k,n_k)$ integers.  The Fourier transform of ${\bf x}(t)$ or 
${\bf v}(t)$
will therefore consist of a set of spikes at discrete frequencies
$\omega_k=l_k\Omega_1 + m_k\Omega_2 + n_k\Omega_3$ that are linear
combinations of the $N$ fundamental frequencies $\Omega_i$, with spectral
amplitudes ${\bf X}_k({\bf J})$ and ${\bf V}_k({\bf J})$.

\section{Trajectory-Following Approaches}
The most straightforward, and probably the most robust, approach to torus
construction is via Fourier analysis of the numerically-integrated
trajectories (Percival 1974; Boozer 1982; Binney \& Spergel 1982, 1984;
Kuo-Petravic et al. 1983; Eaker et al. 1984; Martens \& Ezra 1985).  The
Fourier decomposition of a quasiperiodic orbit (Equation 3) yields a discrete
frequency spectrum.  The precise form of this spectrum depends on the
coordinates in which the orbit is integrated, but certain of its properties
are invariant, including the $N$ fundamental frequencies $\Omega_i$ from
which every line is made up, $\omega_k=l_k\Omega_1 + m_k\Omega_2 +
n_k\Omega_3$.  Typically the strongest line in a spectrum lies at one of the
fundamental frequencies; once the $\Omega_i$ have been identified, the
integer vectors $(l_k,m_k,n_k)$ corresponding to every line $\omega_k$ are
uniquely defined, to within computational uncertainties. Approximations to
the actions may then be computed using Percival's (1974) formulae; e.g. the
action associated with $\theta_1$ in a 3 DOF system is
\begin{equation} J_1 = \sum_k
l_k\left(l_k\Omega_1+m_k\Omega_2+n_k\Omega_3\right) |{\bf X}_k|^2 \label{perc1}
\end{equation} and similarly for $J_2$ and $J_3$, upon replacing the first
factor in the summation by $m_k$ and $n_k$ respectively. Finally, the maps
$(\theta \rightarrow {\bf x})$ are obtained by making the substitution
$\Omega_it\rightarrow\theta_i$ in the  spectrum, e.g.
\begin{eqnarray} x(t) & = &\sum_kX_k(J) \exp\left[ i\left(l_k\Omega_1 +
m_k\Omega_2 + n_k\Omega_3\right)t\right] \nonumber \\ & = &\sum_kX_k(J)
\exp\left[ i\left(l_k\theta_1 + m_k\theta_2 + n_k\theta_3\right)\right]
\nonumber \\ & = & x(\theta_1, \theta_2, \theta_3).
\end{eqnarray} Trajectory following algorithms are easily automated; for
instance, integer programming may be used to recover the vectors
$(l_k,m_k,n_k)$ (Valluri \& Merritt 1998).
 
Binney \& Spergel (1982) pioneered the use of trajectory-following algorithms
for galactic potentials. They integrated orbits for a time $T$ and computed
discrete Fourier transforms, yielding spectra in which each frequency spike
was represented by a peak with finite width $\sim\pi/T$ centered on
$\omega_k$. They then fitted these peaks to the expected functional form
$X_k\sin[(\omega-\omega_k)T]/(\omega-\omega_k)$ using a least-squares
algorithm. They were able to recover the fundamental frequencies in a 2 DOF
potential with an accuracy of $\sim 0.1\%$ after $\sim 25$ orbital periods.
Binney \& Spergel (1984) used equation (4) to construct the ``action map''
for orbits in a principal plane of the triaxial logarithmic potential.
Carpintero \& Aguilar (1998) and Copin, Zhao \& de Zeeuw (this volume)
applied similar algorithms to motion in 2- and 3 DOF potentials.
 
The accuracy of Fourier transform methods can be greatly improved by
multiplying the time series with a windowing function before transforming.
The result is a reduction in the amplitude of the side lobes of each
frequency peak at the expense of a broadening of the peaks; the amplitude
measurements are then effectively decoupled from any errors in the
determination of the frequencies.  Laskar (1988, 1990) developed this idea
into a set of tools, the ``numerical analysis of fundamental frequencies''
(NAFF), which he applied to the analysis of weakly chaotic motion in the
solar system.  Laskar's algorithm recovers the fundamental frequencies with
an error that falls off as $T^{-4}$ (Laskar 1996), compared with $\sim
T^{-1}$ in algorithms like Binney \& Spergel's (1982).  Even for modest
integration times of $\sim 10^2$ orbital periods, the NAFF algorithm is able
to recover fundamental frequencies with accuracies of $\sim10^{-8}$ or better
in many potentials. The result is a very precise representation of the torus
(Figure 2).
 
\begin{figure} \plotfiddle{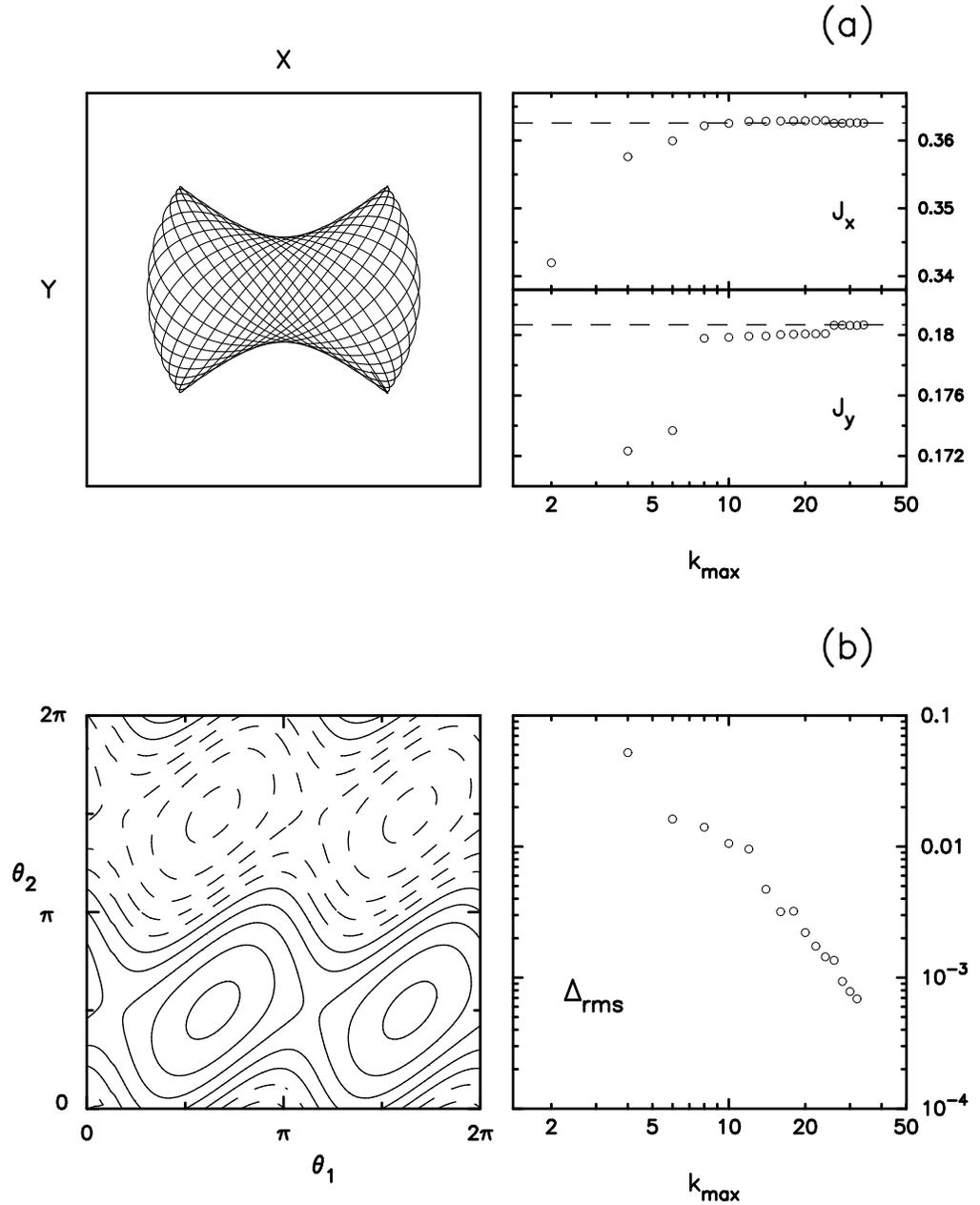}{16cm}{0}{72}{72}{-215}{-30}
\footnotesize
\caption{ Construction of a 2 DOF, box-orbit torus in a St\"ackel potential
using the NAFF trajectory-following algorithm.  (a) The orbit and its
actions, computed using Equation (4) with $k_{max}$ terms. Dashed lines show
the exact $J_i$.  (b) The map $y(\theta_1,\theta_2)$; dashed contours
correspond to negative values of $y$. $\Delta(k_{max})$ is the RMS error in
the reconstructed map, calculated using an equation similar to (5);
$\Delta\sim k_{max}^{-2}$. } \label{fig2}
\end{figure}
 
One drawback of trajectory-following algorithms is the need to extract a
large number of terms in the frequency spectrum in cases where the time
dependence of the integration variables is very different from that of the
angles. This problem may be dealt with by expressing the
numerically-integrated orbit in terms of a set of coordinates that are closer
to the angle variables before computing the Fourier transform; for instance,
tube orbits are most efficiently expressed in the canonically-conjugate
Poincar\'e variables (related to cylindrical coordinates, e.g. Papaphilippou
\& Laskar 1996).
 
Trajectory-following algorithms also suffer from the fundamental limitation
that they must follow the trajectory sufficiently far to see the longest
periodicities of the orbit. In other words, the trajectory must adequately
sample the surface of its invariant torus. Near a resonant torus, i.e. a
torus for which the $\Omega_i$ satisfy a relation
\begin{equation} \sum_{i=1}^N \alpha_i\Omega_i=0
\end{equation} with the $\alpha_i$ integers, trajectories fill their tori
very slowly, necessitating long integration intervals. However even for
near-resonant orbits, one can still efficiently recover the terms in the
spectrum associated with the ``faster'' angles, as well as a reasonable
approximation to the ``slow'' frequency $\Omega_3$ associated with libration
around the resonant orbit (Merritt \& Valluri 1999), and for many
applications these are sufficient. If the precise dependence of the map on
$\theta_3$ is also needed, one possible approach (Papaphilippou \& Laskar
1998) is to integrate the equations of motion using as a time step the period
associated with one of the fast angles, thus eliminating it from the spectrum.
 
Since Fourier techniques focus on the frequency domain, they are particularly 
well suited to identifying regions of phase space occupied by resonances.
They are also ideal for studying the effect of resonances on the structure of
phase space, even in cases where the full tori are difficult to reconstruct.
Resonant tori are places where perturbation expansions of integrable systems
break down, due to the ``problem of small denominators''. In perturbed 
(non-integrable) potentials, one expects stable resonant tori to generate 
regions of regular motion and unstable resonant tori to give rise to chaotic
regions.  Algorithms like NAFF allow one to construct a ``frequency map'' 
of the phase space: a plot of the ratios of the fundamental frequencies
$(\Omega_1/\Omega_3,\Omega_2/\Omega_3)$ for a large a set of orbits selected
from a uniform grid in initial condition space. Resonances appear on the
frequency map as lines, either densely filled lines in the case of stable
resonances, or gaps in the case of unstable resonances; the frequency map
is effectively a representation of the Arnold web (Laskar 1993).
Papaphilippou \& Laskar (1996, 1998),  Wachlin \& Ferraz-Mello (1998) and
Valluri \& Merritt (1998) used frequency maps to study the effect of
resonances on the structure of phase space in triaxial potentials.
 
\begin{figure}
\plotfiddle{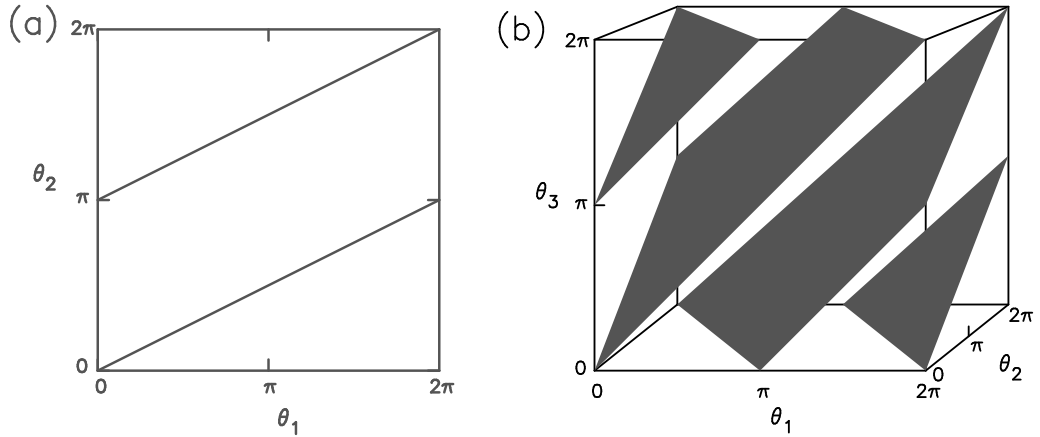}{6cm}{-90}{65}{65}{-280}{+300}
\footnotesize
\caption{Resonant tori. (a) A two-dimensional torus, shown here as a square
with identified edges. The plotted trajectory satisfies a $2:1$ resonance
between the fundamental frequencies, $\Omega_1 - 2\Omega_2 = 0$ (e.g. a
``banana''). (b) A three-dimensional torus, shown here as a cube with
identified sides. The shaded region is covered densely by a resonant
trajectory for which $2\Omega_1 + \Omega_2 - 2\Omega_3 = 0$. This trajectory
is not closed, but it is restricted by the resonance condition to a
two-dimensional subset of the torus. The orbit in configuration space is thin
(Figure 4).}
\end{figure}
 
Precisely resonant orbits can be reconstructed using trajectory-following
algorithms in a particularly straightforward way. A resonance has the effect
of restricting an orbit to a subset of its torus, reducing the number of
independent angle variables by one; thus a 2 DOF trajectory is reduced to a
closed curve and a 3 DOF trajectory becomes a thin sheet (Born 1960;
Goldstein 1980; Figure 3). The two frequencies defining motion on a resonant
three-torus may be taken to be
\begin{equation} \Omega_0^{(1)} = \Omega_3/\alpha_1,\ \ \ \ \Omega_0^{(2)} =
\Omega_2/\alpha_1,
\end{equation} in terms of which
\begin{eqnarray}
\Omega_1 & = & -\alpha_3\Omega_0^{(1)} - \alpha_2\Omega_0^{(2)}, \nonumber \\
\Omega_2 & = & \alpha_1\Omega_0^{(2)}, \nonumber \\ \Omega_3 & = &
\alpha_1\Omega_0^{(1)}.
\end{eqnarray}

\noindent 
This definition is not unique since the orbit is not closed.
\begin{figure}
\plotfiddle{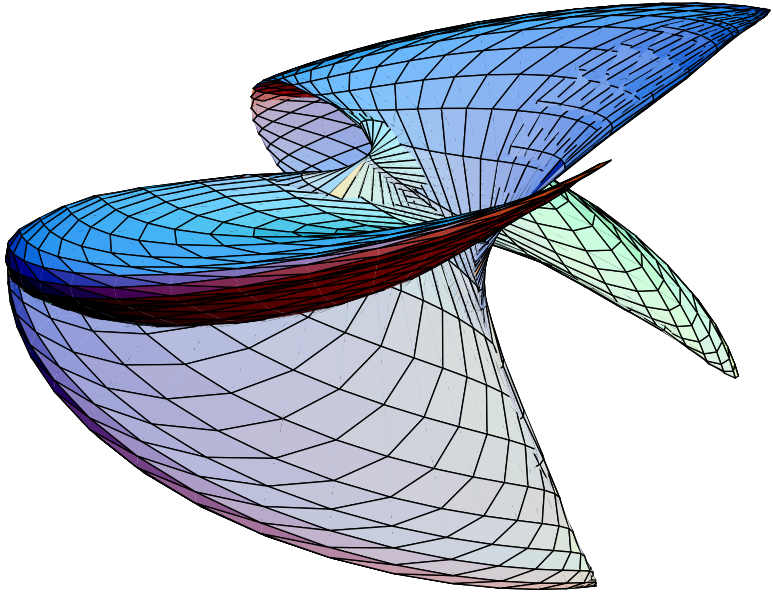}{9cm}{0}{80}{80}{-260}{-180}
\footnotesize
\caption{ The surface in configuration space filled by a resonant, or
``thin,'' box orbit in a triaxial potential (Merritt \& Valluri 1999). The
order of the resonance is $(2,1,-2)$, as in Figure 3b.  The surface was
plotted by representing the spatial coordinates $(x,y,z)$ parametrically
in terms of the two angles that define position on the resonant three-torus.}
\end{figure}
The motion in Cartesian coordinates then becomes
\begin{eqnarray}
{\bf x}(t) & = & \sum_k {\bf X}_k\exp i \left(l_k\Omega_1 + m_k\Omega_2 +
n_k\Omega_3\right)t \nonumber \\
& = & \sum_k {\bf X}_k\exp i \left[\left(-l_k m_3+n_km_1\right)\Omega_0^{(1)} +
\left(-l_k m_2 + m_k m_1\right) \Omega_0^{(2)}\right]t \nonumber \\ & = &
\sum_k {\bf X}_k\exp i \left({l_k}'\Omega_0^{(1)} + {m_k}'\Omega_0^{(2)}\right)t
\nonumber \\ & = & \sum_k {\bf X}_k\exp i \left({l_k}'\theta^{(1)} +
{m_k}'\theta^{(2)}\right) \nonumber \\ & = & {\bf x}(\theta^{(1)}, \theta^{(2)}).
\end{eqnarray} The result is a set of parametric expressions for the
Cartesian coordinates $(x,y,z)$ in terms of the angles $(\theta^{(1)},
\theta^{(2)})$ that define position on the two-torus (Figure 4).
 
\section{Iterative Approaches}
 
Iterative approaches to torus construction consist of finding successively
better approximations to the map $\theta \rightarrow {\bf x}$ given some initial
guess ${\bf x}(\theta)$; canonical perturbation theory is a special case, and in
fact iterative schemes often reduce to perturbative methods in appropriate
limits. Iterative algorithms were first developed in the context of
semi-classical quantization for computing energy levels of bound molecular
systems, and they are still best suited to assigning energies to actions,
$H({\bf J})$. Most of the other quantities of interest to galactic dynamicists --
e.g. the fundamental frequencies $\Omega_i$ -- are not recovered with high
accuracy by these algorithms. Iterative schemes also tend to be numerically
unstable unless the initial guess is close to the true solution. On the other
hand, iterative algorithms can be more efficient than trajectory-following
methods for orbits that are near resonance.
 
Ratcliff, Chang \& Schwarzschild (1984) pioneered iterative schemes in
galactic dynamics. They noted that the equations of motion of a 2 DOF regular
orbit,
\begin{equation} \ddot x = -{\partial\Phi\over\partial x},\ \ \ \ \ddot y =
-{\partial\Phi\over\partial y},
\end{equation} can be written in the form
\begin{eqnarray} \left(\Omega_1{\partial\over\partial\theta_1} +
	\Omega_2{\partial\over\partial\theta_2}\right)^2x & = &
	-{\partial\Phi\over\partial x}, \nonumber \\
\left(\Omega_1{\partial\over\partial\theta_1} +
	\Omega_2{\partial\over\partial\theta_2}\right)^2y & = &
	-{\partial\Phi\over\partial y}.
\end{eqnarray}
If one specifies $\Omega_1$ and $\Omega_2$ and treats $\partial\Phi/\partial
x$ and $\partial\Phi/\partial y$ as functions of the $\theta_i$, equations
(11) can be viewed as nonlinear differential equations for
$x(\theta_1,\theta_2)$ and $y(\theta_1,\theta_2)$. Ratcliff et al. expressed
the coordinates as Fourier series in the angle variables,
\begin{equation}
{\bf x}(\theta) = \sum_{n} {\bf X}_{n} e^{i{n}\cdot{\theta}}.
\end{equation}
Substituting (12) into (11) gives
\begin{equation} \sum_{n}({\bf n}\cdot\Omega)^2{\bf X}_{n}e^{i{n \cdot
\theta}}=\nabla\Phi
\end{equation}
where the right hand side is again understood to be a function of the angles.
Ratcliff et al. truncated the Fourier series after a finite number of terms
and required equations (13) to be satisfied on a grid of points around the
torus. They then solved for the ${\bf X}_n$ by iterating from an initial guess.
Convergence was found to be possible if the initial guess was close to the
exact solution. A similar algorithm was developed for recovering tori in the
case that the actions, rather than the frequencies, are specified a priori.
Guerra \& Ratcliff (1990) applied these algorithms to motion in the plane of
rotation of a nonaxisymmetric potential.
 
Another iterative approach to torus construction was developed by Chapman,
Garrett \& Miller (1976) in the context of semiclassical quantum theory. One
begins by dividing the Hamiltonian $H$ into separable and non-separable parts
$H_0$ and $H_1$, then seeks a generating function $S$ that maps the known
tori of $H_0$ into tori of $H$. For a generating function of the $F_2$-type
(Goldstein 1980), one has
\begin{equation}
{\bf J}(\theta,{\bf J}') = {\partial S\over\partial\theta}, \ \ \ \
\theta'(\theta,{\bf J}') = {\partial S\over\partial {\bf J}'}
\end{equation}
where $({\bf J}, \theta)$ and $({\bf J}', \theta')$ are the action-angle variables
of $H_0$ and $H$ respectively. The generator $S$ is determined, for a
specified ${\bf J}'$, by substituting the first of equations (14) into the
Hamiltonian and requiring the result to be independent of
$\theta$. One then arrives at $H({\bf J}')$. Chapman et al. showed that a
sufficiently general form for $S$ is
\begin{equation}
S(\theta,{\bf J}') = \theta \cdot {\bf J}' - i\sum_{{n}\ne 0}
S_{n}({\bf J}')e^{i{n}\cdot\theta},
\end{equation}
where the first term is the identity transformation, and they evaluated a
number of iterative schemes for finding the $S_{n}$. One such scheme was
found to recover the results of first-order perturbation theory after a
single iteration. McGill \& Binney (1990) refined the Chapman et al.
algorithm and applied it to 2 DOF motion in the axisymmetric logarithmic
potential.
 
The generating function approach is not naturally suited to deriving the
other quantities of interest to galactic dynamicists. For instance, equation
(14) gives $\theta'(\theta)$ as a derivative of $S$, but since $S$ must be
computed separately for every ${\bf J}'$ its derivative is likely to be
ill-conditioned. Binney \& Kumar (1993) and Kaasalainen \& Binney (1994a)
discussed two schemes for finding $\theta'(\theta)$; the first requires the
solution of a formally infinite set of equations, while the latter requires
multiple integrations of the equations of motion for each torus --
effectively a trajectory-following scheme.
 
Kaasalainen \& Binney (1994a) noted that the success of the generating
function method depends strongly on the choice of $H_0$. For box orbits,
which are most naturally described as coupled rectilinear oscillators, they
found that a harmonic-oscillator $H_0$ gave poor results unless an additional
point transformation was used to deform the rectangular orbits of $H_0$ into
narrow-waisted boxes like those in typical galactic potentials. Kaasalainen
(1995a) considered orbits belonging to higher-order resonant families and
found that it was generally necessary to define a new coordinate
transformation for each family.
 
Warnock (1991) presented a hybrid scheme in which the generating function $S$
was derived by numerically integrating an orbit from appropriate initial
conditions, transforming the coordinates to $({\bf J}, \theta)$ of $H_0$ and
interpolating ${\bf J}$ on a regular grid in $\theta$. The values of the
$S_{n}$ then follow from the first equation of (14) after a discrete
Fourier transform. Kaasalainen \& Binney (1994b) found that Warnock's scheme
could be used to substantially refine the solutions found via their iterative
algorithm. Another hybrid scheme was discussed by Reiman \& Pomphrey (1991).
 
Having computed the energy on a grid of ${\bf J}'$ values, one can interpolate to
obtain the full Hamiltonian $H({\bf J}')$. If the system is not in fact
completely integrable, this $H$ may be rigorously interpreted as smooth
approximation to the true $H$ (Warnock \& Ruth 1991, 1992) and can be taken
as the starting point for secular perturbation theory. Kaasalainen (1994)
developed this idea and showed how to recover accurate surfaces of section in
the neighborhood of low-order resonances in the planar logarithmic potential.
 
Percival (1977) described a variational principle for constructing tori. His
technique has apparently not yet been implemented in the context of galactic
dynamics.
 
\section{Chaotic Motion}
 
Torus-construction machinery may be applied to orbits that are approximately,
but not precisely, regular (Laskar 1993).  The frequency spectrum of a weakly
chaotic orbit will typically be close to that of a regular orbit, with most
of the lines well approximated as linear combinations of three ``fundamental
frequencies'' $\Omega_i$. However these frequencies will change with time as
the orbit diffuses from one ``torus'' to another.  The diffusion rate can be
measured via quantities like $|\Omega_1 - \Omega'_1|$, the change in a
``fundamental frequency'' over two consecutive integration intervals.
Papaphilippou \& Laskar (1996, 1998), Valluri \& Merritt (1998) and Wachlin
\& Ferraz-Mello (1998) used this technique to study chaos and diffusion in
triaxial galactic potentials. Kaasalainen (1995b) showed that approximate
tori could be constructed even in chaotic phase space via the hybrid scheme
of Warnock (1991). While such tori clearly do not describe the motion of
chaotic orbits over long times, they are useful for understanding the onset
of chaos and its relationship to resonances, as well as for studying
evolution of the phase-space distribution function in action space via the
Fokker-Plank equation (Lichtenberg \& Leiberman 1992).
 
\section{Summary}
 
Trajectory-following schemes for torus construction are robust and easily
automated. They can recover the fundamental frequencies with great precision
and are well suited to studies of weak chaos and for mapping resonances.
However they are inefficient for constructing the full torus of an orbit that
lies close to, but slightly off of, a resonance. Iterative techniques are
efficient for assigning energies to actions but less suited to recovering the
other quantities of interest to galactic dynamicists, such as the fundamental
frequencies. However they can be more efficient than trajectory-following
algorithms for constructing nearly-resonant tori. Hybrid schemes that combine
features of both approaches show considerable promise.

\end{document}